\documentclass[useAMS,usenatbib]{mn2e}
\usepackage{newtxtext,newtxmath} 
\usepackage{epsfig,comment}
\usepackage{color}
\usepackage{natbib}
\usepackage{deluxetable}
\usepackage{amsmath}
\usepackage{graphicx}
\usepackage[colorlinks=true,citecolor=blue]{hyperref}
\usepackage{epstopdf}
\usepackage{gensymb}
\usepackage{amsmath}
\usepackage[dvipsnames]{xcolor}

\def\kms{km s$^{-1}$}

\twocolumn
\title[Spectro-polarimetry of 1H0323+342]{Spectro-polarimetric view of the gamma-ray emitting NLS1 1H0323+342}

\author[Jose et al.]{Jincen Jose$^{1,2}$, Suvendu Rakshit$^{1}$\thanks{E-mail: suvenduat@gmail.com}, Swayamtrupta Panda$^{3,4,\thanks{CNPq Fellow}, \thanks{NSF/Gemini Science Fellow}}$, Jong-Hak Woo$^{5}$, \newauthor  C. S. Stalin$^{6}$, Neha Sharma$^{1}$ and Shivangi Pandey$^{1,7}$ \\ 
 \\$^{1}$ Aryabhatta  Research Institute of Observational Sciences (ARIES), Manora Peak, Nainital, 263002 India \\
 $^{2}$ Center for Basic Sciences, Pt. Ravishankar Shukla University, Raipur, Chhattisgarh, 492010, India \\
 $^{3}$ Laborat\'orio Nacional de Astrof\'isica - MCTI, R. dos Estados Unidos, 154 - Na\c{c}\~oes, Itajub\'a - MG, 37504-364, Brazil \\
 $^{4}$ International Gemini Observatory/NSF NOIRLab, Casilla 603, La Serena, Chile \\
 $^{5}$ Department of Physics \& Astronomy, Seoul National University, Seoul 08826, Republic of Korea \\
 $^{6}$ Indian Institute of Astrophysics, Block II, Koramangala, Bangalore 560034, India \\
 $^{7}$ Department of Applied Physics/Physics, M.J.P. Rohilkhand University, Bareilly, Uttar Pradesh - 24300}

\begin{document}
\date{Accepted ---. Received ---; in original form ---}

\pagerange{\pageref{firstpage}--\pageref{lastpage}} \pubyear{2020}

\maketitle
\label{firstpage}


\begin{abstract}
The gamma-ray emitting narrow-line Seyfert 1 galaxies are a unique class of objects that launch powerful jets from relatively lower-mass black hole systems compared to the Blazars. However, the black hole masses estimated from the total flux spectrum suffer from the projection effect, making the mass measurement highly uncertain. The polarized spectrum provides a unique view of the central engine through scattered light. We performed spectro-polarimetric observations of the gamma-ray emitting narrow-line Seyfert 1 galaxy 1H0323+342 using \textit{SPOL/MMT}. The degree of polarization and polarization angle are 0.122 $\pm$ 0.040\% and 142 $\pm$ 9 degrees, while the H$\alpha$ line is polarized at 0.265  $\pm$ 0.280\%. We decomposed the total flux spectrum and estimated broad H$\alpha$ FWHM of 1015 \kms{}. The polarized flux spectrum shows a broadening similar to the total flux spectrum, with a broadening ratio of 1.22. The Monte Carlo radiative transfer code `STOKES' applied to the data provides the best fit for a small viewing angle of 9-24 degrees and a small optical depth ratio between the polar and the equatorial scatters. A thick BLR with significant scale height can explain a similar broadening of the polarized spectrum compared to the total flux spectrum with a small viewing angle.  
\end{abstract}

\begin{keywords}
{\em Methods:} data analysis - {\em Galaxies:} active - {\em Techniques:} polarimetric - {\em Techniques: spectroscopic} - radiative transfer 
\end{keywords}

\section{Introduction}
\label{s:intro}
Narrow-line Seyfert galaxies (NLS1) are the high-accreting active galactic nuclei (AGN) powered by black holes in the mass range of 1-100 million solar mass and characterized by the H$\beta$ emission line width less than 2000 \kms{} and [OIII]$\lambda$5007 to H$\beta$ flux ratio less than 3 \citep{1985ApJ...297..166O}. They show a steep soft X-ray spectrum \citep{2023A&A...669A..37G} and low amplitude of optical variation \citep{2017ApJ...842...96R}. A small fraction of $\sim$7\% are known to be radio-detected, and only a handful of them are detected in high energy $\gamma$-ray \citep{2009ApJ...707L.142A,2019ApJ...872..169P}. Detection of NLS1 in $\gamma$-ray suggests the presence of relativistic jets in relatively lower mass AGN compared to Blazars. Therefore, they are ideal for studying how jets form in the low-mass AGN. However, a few recent studies suggest the black hole masses in these objects are underestimated due to the projection effect \citep{2008MNRAS.386L..15D,2016MNRAS.458L..69B,2019ApJ...881L..24V}.

Spectro-polarimetry is a crucial tool for studying the central engine of AGN by scattering light from the surrounding medium. Long ago, broad emission lines were detected in polarized light in Type 2 AGNs \citep{1985ApJ...297..621A}, leading to the discovery of the well-known unification model of AGNs \citep{1993ARA&A..31..473A, 1995PASP..107..803U}. The polarization properties of type 1 AGNs suggest the presence of scattering material in the equatorial region and the rotation of the polarization angle within the broad lines; however, the polar scattering is also present in some AGNs \citep{2002MNRAS.335..773S, 2005MNRAS.359..846S, 2023A&A...678A..63S}. A few attempts have been made to study radio quiet NLS1s using spectro-polarimetry \citep[e.g.,][]{1989ApJ...342..224G, 1999ApJ...518..219K, 2023A&A...678A..63S}, but none of them showed a significant increase in the H$\alpha$ in polarized light. However, the spectro-polarimetric study of PKS 2004-447, a RL-NLS1, by \citet{2016MNRAS.458L..69B} shows a six times broader H$\alpha$ line in the polarized spectrum than the width seen in the direct light. The wavelength-dependent polarization position angle provides an independent black hole mass, which is in good agreement with those based on the reverberation method \citep[e.g.,][]{2015ApJ...800L..35A}. The black hole masses have been successfully measured for a few objects based on the wavelength-dependent polarization angle variation using the H$\alpha$ line \citep[e.g.,][]{2015MNRAS.448.2879A,2019MNRAS.482.4985A} and Mg II lines \citep{2020MNRAS.497.3047S}.  \citet{2021MNRAS.502.5086C} performed spectro-polarimetry of 25 low-redshift AGNs using FORS2/VLT and found that the black hole mass could be underestimated by a factor of 5 due to the inclination angle.

1H 0323+342 is one of the most interesting $\gamma$-ray-detected NLS1 \citep{2009ApJ...707L.142A} located at a redshift of 0.062 having a H$\beta$ line width of $\sim$ 1600 \kms{} \citep{1993AJ....105.2079R}. High energy $\gamma$-ray detection by the \textit{Fermi} Gamma-ray Space Telescope suggests the object is close to face-on and processes a relativistic jet. The host-galaxy morphology is complex, with indications of a one-armed spiral host structure or a ring-like structure in the optical images \citep{2007ApJ...658L..13Z,2008A&A...490..583A,2014ApJ...795...58L}, which is contrary to the elliptical host found in Blazars having powerful jets \citep{2000ApJ...544..258S}. \citet{2016ApJ...824..149W} have performed reverberation mapping monitoring and estimated a black hole mass of $\log  M_{\mathrm{BH}} = 7.53$, which is one-order less massive than that is expected from the relation between black mass and bulge luminosity \citep{2014ApJ...795...58L}. However, \citet{2016ApJ...824..149W} used virial factor ($f_{\mathrm{BLR}}$) of 6.17 from \citet{2014MNRAS.445.3073P} to estimate the reverberation mapped black hole mass. The $f_{\mathrm{BLR}}$ is highly sensitive to the inclination \citep{2014MNRAS.445.3073P,2015MNRAS.447.2420R, 2019ApJ...882...79P}. Therefore, spectro-polarimetric observation could provide new insights into the BLR geometry and measure independent black hole masses from polarized emission. 

To study the geometry of the broad line region and estimate an independent black hole mass of 1H0323+342, we present the results of our spectro-polarimetric observations using SPOL/MMT in this paper. The structure of the paper is as follows: in section \ref{sec:data}, we describe the observation and data reduction method. The results are presented in section \ref{sec:results}. Modeling of polarized emission is performed in section \ref{sec:modeling}. The results have been discussed in section \ref{sec:discussion} with a summary in section \ref{sec:summary}. Throughout this paper, we assume a $\Lambda$-cold dark matter cosmology with a Hubble constant of H$_{\rm 0}$ = 70 \kms{} Mpc$^{-1}$, $\Omega_{\Lambda}$ = 0.7, and $\Omega_{\rm m}$ = 0.3.

\section{Observations and data reduction}
\label{sec:data}
We observed 1H0323+342 using SPOL \citep{1992ApJ...398L..57S} mounted on 6.5m \textit{MMT} on 20th December 2019. Observations were carried out using a slit of 1.5 arcsec, and the 964 l/mm grating, providing a spectral resolution of
$\sim$15 \AA\, in the wavelength range of 4100–7200\AA\, to cover both the H$\alpha$ and H$\beta$ regions. 
A rotatable semi-achromatic half-wave plate was employed to modulate the incoming polarization. At the same time, a Wollaston prism, having very high throughput and polarimetric efficiency, in the collimated beam separates the orthogonally polarized spectra onto a thinned and anti-reflection-coated 800 × 1200 SITe CCD. Four separate exposures, capturing 16 orientations of the wave plate, were conducted. In each orientation, the ratio of fluxes of the two orthogonally polarized spectra was used to calculate the polarization measurement. This provided us with background-subtracted measures of each normalized linear Stokes parameter, q and u \citep[see][]{1992ApJ...398L..57S,2003ApJ...593..676S}. In total, 1 hour of on-source observation was made to obtain high signal-to-noise ratio (S/N) data.

Unpolarized standard BD+284211 was observed to determine the instrumental polarization, which is very small ($<$0.01\%) in SPOL. Three interstellar polarized standard stars were observed, namely BD+59389, VI Cyg 12, and Hiltner 960. For calibrations, dome flats were taken. Wavelength calibrations were done using He, Ne, and Ar lamp spectra. Standard data reduction is performed using {\sc IRAF}, subtracting the bias using the overscan region with detailed processes outlined in \citet{2003ApJ...593..676S}.


The degree of polarization (P) and position angle ($\chi$) were calculated using normalized linear polarization Stokes parameters q and u. 
\begin{equation}
\label{Eq:PD}
    P = \sqrt{( q^2 + u ^2)} 
\end{equation}
\begin{equation}
\label{Eq:PA}
    \chi = 0.5 \times \arctan(u/q) 
\end{equation}

Following \citet{1974ApJ...194..249W}, we attempted debiasing the degree of polarization \citep[see also Eq. 1 of][]{2009arXiv0912.3621S} to estimate the statistical bias inherent in these measurements.  However, the correction was negligible due to binning the spectra (see section \ref{s:pol}). We found that debiasing changes the results of the degree of polarization by $<0.02\%$ only, which is insignificant. Therefore, this polarization debiasing is not considered for further analysis. The polarization angles are adjusted to ensure they fall within the range of 0 to 180 degrees, using the formula and conditions provided by \cite{Bagnulo_2009}.

\begin{figure}
\centering
\includegraphics[height=7cm,width=9cm]{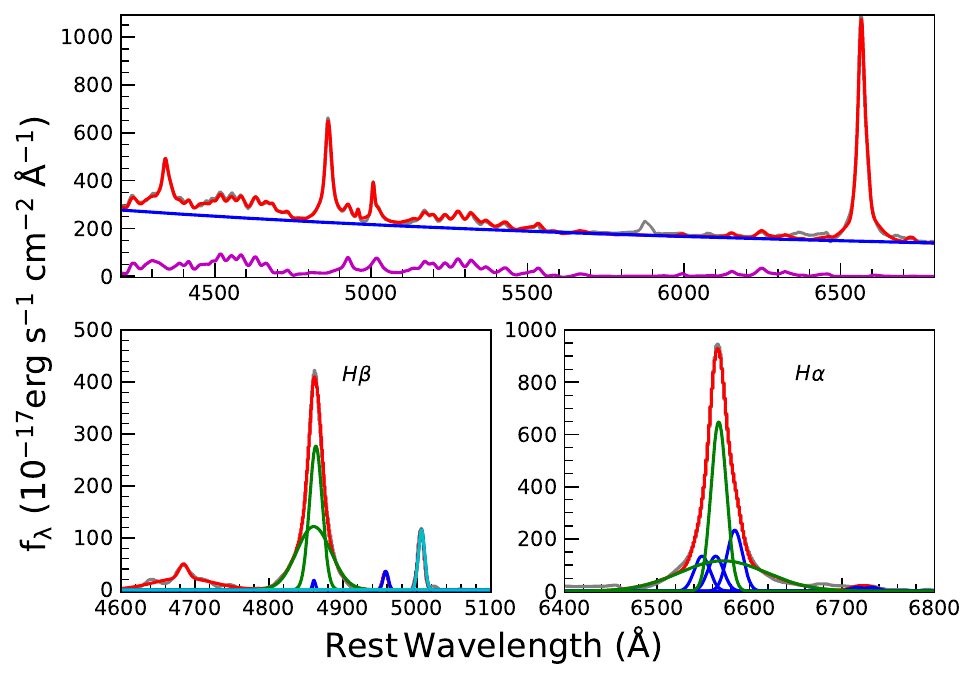}
\caption{Spectral decomposition of the total flux spectrum. The observed data (grey), best-fit model (red), power-law (blue), and decomposed Fe II emission (magenta) are shown in the top panel. The bottom panel shows the fitting in the H$\beta$ and H$\alpha$ regions. The broad components are shown in green, while the narrow components are shown in blue.}
\label{Fig:fig_spec}
\end{figure}

\begin{table}
    \centering
    \begin{tabular}{l|c|r} \hline \hline
    Quantities & Units &  Values \\ \hline
$\log L_{5100}$ & erg s$^{-1}$       & 43.9 $\pm$ 0.1 \\
PL index &  --  & -1.43 $\pm$ 0.01 \\
FWHM (H$\beta$ BC) & \kms{}   & 1554 $\pm$ 9 \\
$\log L$ (H$\beta$ BC) & erg s$^{-1}$  & 41.98 $\pm$ 0.01 \\
$\log L$ (H$\beta$ NC) & erg s$^{-1}$  & 39.79 $\pm$ 0.01 \\
$\log L$ ([OIII]$\lambda$5007) & erg s$^{-1}$ & 40.95 $\pm$ 0.01 \\
FWHM (H$\alpha$ BR) & \kms{} & 1015 $\pm$ 1 \\
$\log L$ (H$\alpha$ BC) & erg s$^{-1}$  & 42.31 $\pm$ 0.01\\
$\log L$ (H$\alpha$ NC) & erg s$^{-1}$  & 41.31 $\pm$ 0.01 \\
$\log L$ ([NII]$\lambda$6585) & erg s$^{-1}$ & 41.55 $\pm$ 0.01 \\
R$_{4570}$ &   --    & 1.14 $\pm$ 0.01  \\
$\log M_{\rm BH}$ & M$_{\odot}$    & 7.24 $\pm$ 0.01 \\ \hline \hline
    \end{tabular}
    \caption{Spectral properties from the decomposition of the total flux spectra.}
    \label{tab:spectral_decomposition}
\end{table}

\begin{figure*}
\centering
\includegraphics[height=10cm,width=14cm]{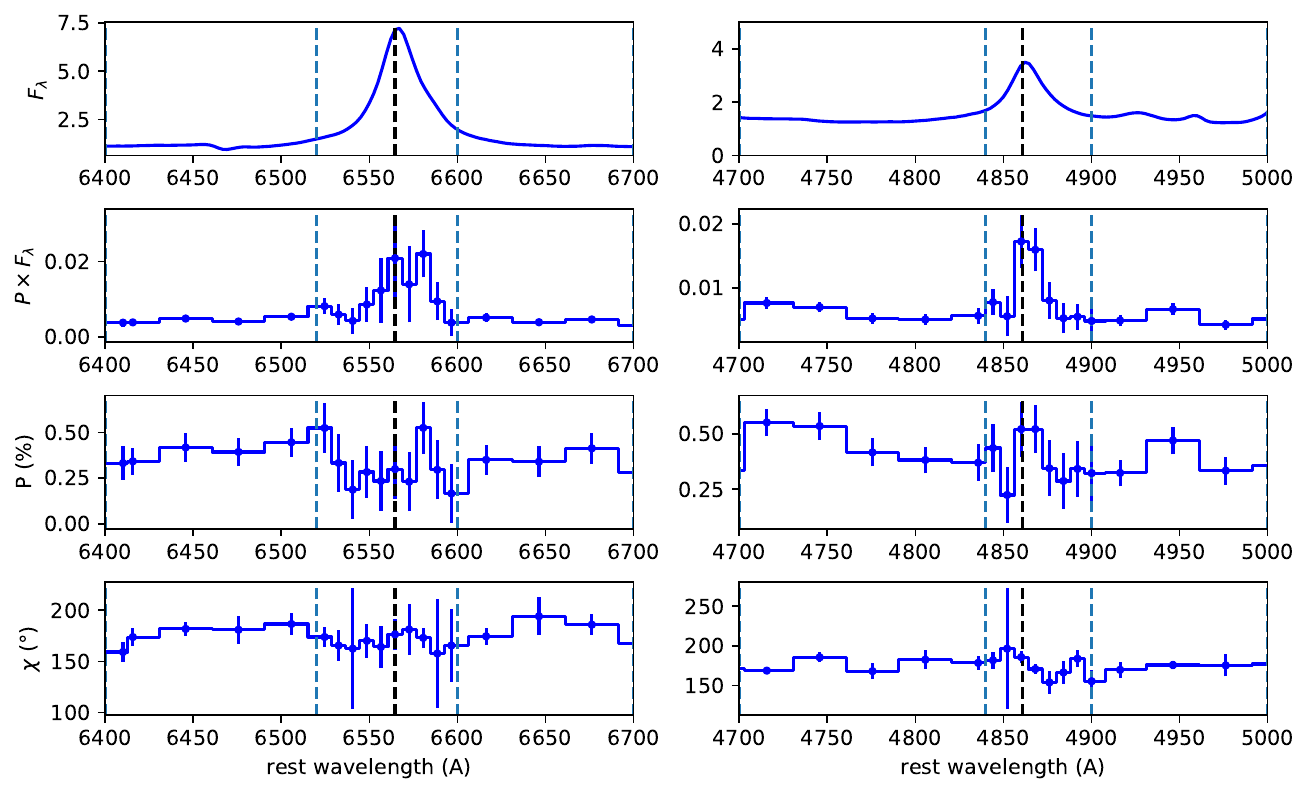}
\caption{The total flux spectrum (top panel), polarized flux (2nd panel), degree of polarization (3rd panel), and polarization position angle (4th panels) are shown for H$\alpha$ and H$\beta$ region. The region within the vertical lines is binned by 8 \AA, while the outside region is binned by 30 \AA.}
\label{Fig:fig_pol}
\end{figure*}

\section{Results}\label{sec:results}

\subsection{Total flux spectra}\label{sec:flux}
Figure \ref{Fig:fig_spec} shows the total flux spectrum. To estimate the continuum and emission line properties, we performed spectral decomposition of the total flux spectrum described in \citet{2020ApJS..249...17R} using the publicly available code \textsc{PyQSOFit}\footnote{\url{https://github.com/legolason/PyQSOFit}} 
developed by \citet{2018ascl.soft09008G}. First, we corrected the spectrum for galactic extinction using the \citet{1998ApJ...500..525S} map and the Milky Way extinction law of \citet{1999PASP..111...63F} with $R_V=3.1$. We performed multi-component modeling of the continuum and emission lines. Second, we model the continuum using a combination of power-law in the form of $f_{\lambda}=\beta \times \lambda^{\alpha}$ and the optical Fe II template from \citet{1992ApJS...80..109B}.    

The best-fit continuum model was subtracted from the spectrum, and the residual spectrum containing emission lines was modeled. The H$\beta$ and H$\alpha$ regions were modeled using multiple Gaussians. All the narrow components were modeled using a single Gaussian except [OIII]$\lambda\lambda$4959,5007  doublets, which were modeled using double Gaussians each. The broad components of H$\beta$ and H$\alpha$ were modeled using two Gaussians. The H$\alpha$ line complex was modeled with $H\alpha$ broad and narrow components, [NII]$\lambda\lambda$6549,6585 and [SII]$\lambda\lambda$6718,6732. The flux ratio of [SII] doublets was fixed to unity, while [NII] and [OIII] doublets were fixed to their theoretical values. All the lines within a given complex were fitted simultaneously. Uncertainty in the parameter was estimated by Monte Carlo simulations, creating mock spectra where original flux values are modified adding Gaussian random deviates of zero mean and standard deviation given by the flux uncertainty. Uncertainty quoted here is the 16$^{\rm th}$ and 84$^{\rm th}$ percentile values of the distribution of each quantity.  

The optical spectrum shows narrow Balmer lines of FWHM 1554 \kms{} for H$\beta$ and 1015 \kms{} for H$\alpha$, consistent with the previous observations \citep{2016ApJ...824..149W}. The Fe II strength ($R_{4570}$), the ratio of Fe II (4434-4684) to H$\beta$ \citep{1992ApJS...80..109B}, is 1.14, suggesting a very strong Fe II emission in this object. We estimated a black hole mass of $\log M_{\mathrm{BH}}$ [M$_{\odot}$] = 7.24, using the virial relation given by \citet{Woo2015} with a scale factor \textit{f}=1.12 and FWHM of the H$\beta$ line. We note that the $f$ is a major source of uncertainty in the single-epoch black hole mass estimates based on virial relation. Here, we used $f=1.12$ obtained by \citet{Woo2015} based on a narrow-line Seyfert 1 galaxies sample in which category 1H0323+342 belongs. However, \citet{2016ApJ...824..149W} have used higher $f$ values from \citet{2014MNRAS.445.3073P}, considering that the source is viewed face-on, but as we discussed in section \ref{sec:geometry}, 1H0323+342 has a large disk opening angle and $f \sim 1$ as it weakly depends on inclination. Thus, it justifies using $f=1.12$ for the black hole mass estimation from the flux spectrum.. We highlight the salient spectral properties obtained from our spectral decomposition and fitting in Table \ref{tab:spectral_decomposition}.

\begin{table*}
\begin{center}
\begin{tabular}{llllllll}
\hline \hline
Name & RA & DEC & Properties & $P_{\mathrm{mean}}$& $\chi_{\mathrm{mean}}$& $q_{\mathrm{mean}}$& $u_{\mathrm{mean}}$\\
     &   &      &            &  (\%)  &  (\degree) & (\%) & (\%) \\\hline
BD +28 4211 &  21 51 11.02  &  +28 51 50.36 & Unpolarized  & 0.092 $\pm$ 0.007 & 140 $\pm$ 2 & -0.091 $\pm$ 0.007 &  0.017 $\pm$ 0.006 \\ \hline
BD+59 389 & 02 02 42.09  & +60 15 26.44    & Polarized &  6.645 $\pm$ 0.006 &  97 $\pm$ 1 &  -1.787 $\pm$ 0.006 &  -6.401 $\pm$ 0.007 \\
VI Cyg 12    & 20 32 41.10  & +41 14 28.00  & Polarized  & 8.874 $\pm$ 0.034 &  116 $\pm$ 1 &  -7.009 $\pm$ 0.035 &  -5.443 $\pm$ 0.034 \\
Hiltner 960 & 20 23 28.60  & +39 20 57.00 & Polarized & 5.621 $\pm$ 0.009  & 54 $\pm$ 1 &  5.327 $\pm$ 0.009 &  -1.795 $\pm$ 0.009 \\ \hline \hline
\end{tabular}
\caption{The calculated value of the degree of polarization and polarization angle ($\chi$) measured for the whole wavelength range, continuum windows, and H$\alpha$ emission line.}
\label{tab:PD_PA_std}
\end{center}
\end{table*}

\begin{table*}
\begin{center}
\begin{tabular}{lllllll}
\hline
Name & $P_{\mathrm{mean}}$ & $\chi_{\mathrm{mean}}$   & $P_{\mathrm{cont}}$  & $\chi_{\mathrm{cont}}$  & $P_{\mathrm{line}}$  & $\chi_{\mathrm{line}}$    \\ 
 &  (\%)    & (\degree) & (\%) & (\degree) & (\%) & (\degree)\\ \hline
1H0323+342 & 0.166 $\pm$ 0.007 (0.354 $\pm$ 0.007)     & 173  $\pm$  1 (158 $\pm$  1)  & 0.122 $\pm$ 0.040 (0.344 $\pm$ 0.040)  & 142 $\pm$ 9  (146 $\pm$ 3) & 0.265  $\pm$ 0.280   & 181  $\pm$ 52          \\ \hline
\end{tabular}
\caption{The calculated mean value of the degree of polarization and polarization angle ($\chi$) measured for the whole wavelength range (including both line and continuum), only continuum windows, and H$\alpha$ emission line for 1H0323+342. The values in the parenthesis are obtained after ISP correction.}
\label{tab:PD_PA}

\end{center}
\end{table*}

\subsection{Polarized emission}
\label{s:pol}
To obtain information about the  polarized emission, we calculated the mean polarization, where mean Q and mean U are used\footnote{here, Q = \textit{q*I}, and U = \textit{u*I}, where \textit{q} and \textit{u} are normalized linear polarization Stokes parameters, and \textit{I} is the total flux.} to calculate the degree of polarization (Eq: \ref{Eq:PD}) 
and polarization angle (Eq: \ref{Eq:PA})
for the whole wavelength range of the spectrum. Table \ref{tab:PD_PA_std} summarizes the mean polarization obtained for one unpolarized standard and three polarized standard stars. The degree of polarization and position angles are consistent with the literature value\footnote{\url{http://james.as.arizona.edu/~psmith/SPOL/polstds.html}}. For example, the degree of polarization is found to be $6.645 \pm 0.006$\%, $8.874 \pm 0.034$\%, and $5.621 \pm 0.009$\% for the three polarized stars, namely, BD+59 389, VI Cyg 12 and Hiltner 960, respectively (see Table \ref{tab:PD_PA_std}), which agree well with the literature values of 6.3 (B) to 6.7 (V), 8.9 (V) to 7.8 (R) and 5.7 (B) to 5.2 (R) as provided in SPOL website. The degree of polarization for the unpolarized standard star is estimated at $0.092\pm0.007\%$.  

Figure \ref{Fig:fig_pol} shows the total flux, the polarized flux, and the polarization degree for 1H0323+342. Using the above-mentioned method, we calculated the mean continuum polarization for 1H0323+342. Furthermore, we calculated the mean Q and mean U, taking the region of nearly 200 \AA, flanked by the width of 100 \AA, considering 6564.61 \AA\, as the central value, and then calculated continuum polarization as above. The measured values are reported in the table \ref{tab:PD_PA}. Note that the mean polarization can be calculated by method-1) taking the mean Q and mean U and then calculating P, or method-2) first calculating the P and then taking the mean. We used the first method as only positive signs are considered in flux. Squaring all values along the array to calculate the mean P may artificially increase the polarization degree. If we use the second method, the mean degree of polarization over the entire spectrum is increased to 0.38\% from the 0.16\% (see Table \ref{tab:PD_PA}) obtained in the first method. Similarly, the $P_{\mathrm{cont}}$ also increased to 0.43\% from 0.12\%.

\subsubsection{Effect of Inter-stellar polarization}

The Interstellar Medium (ISM) significantly contributes to the measured polarization. This is primarily due to the dust scattering along the line of sight, governed by the total dust column density. Therefore, the total polarization is $Q_m$, $U_{m}$ =  $Q_{ISM}$, $U_{ISM}$ + $Q_{AGN}$, $U_{AGN}$. It is necessary to properly subtract the ISM polarization for a source with low polarization, such as this. Unfortunately, there is no straightforward way to correct for the ISM polarization. \citet{2021MNRAS.508...79J} used three low-polarization stars in the same direction as their source, Fairall 9. They measured their polarization, assuming that their observed polarization was entirely caused by ISM scattering. However, due to the lack of such observations of stars towards 1H0323+342, we can not apply such a method to correct ISM polarization. Only unpolarized star BD+284211 was observed, which is very far from the source. Galactic extinction in the direction of the source could also provide some information about the possible ISM polarization. The Galactic extinction of 1H0323+342, measured from NED, is $A_{V}=0.57$, which is quite high, suggesting Interstellar polarization is non-negligible. 

\citet{2021MNRAS.502.5086C} corrected the interstellar polarization for their sources based on the \textit{Planck} polarization images. Recently, this method has been described in detail for IRAS 04416+1215 (A$_{\rm V}$=1.187) by \citet{2023A&A...678A..63S}. We followed the same method and calculated interstellar polarization based on I, Q, and U \textit{Planck} images at frequency 353 GHz from skyview\footnote{\url{https://skyview.gsfc.nasa.gov/current/cgi/query.pl}}. The images were re-sampled, smoothing by 9 pixels. The submillimeter polarization vector ($q_s$, $u_s$) is converted to the visible using $q_V =-(q_S \tau_V)/R_{(S/V)}$ and $u_V =-(u_S \tau_V)/R_{(S/V)}$ where the ratio of submillimeter to V-band polarization $R_{(S/V)}$ = $4.2 \pm 0.2$ (stat.) $\pm$ 0.3 (syst.), fairly constant across different lines of sight in the diffuse interstellar medium \citep[see][]{2015A&A...576A.106P}, and $\tau_{V}$ is the absorption optical depth. Then, the empirical law from \citet{1975ApJ...196..261S} is applied to convert the V-band ISP polarization to the wavelength of the H$\alpha$ line using 5500$\AA$ as the reference point at which the ISP reaches its maximum ($\lambda_{\mathrm{max}}$). The calculated ratio between $p_{(H\alpha)}$ and $p_V$, is $\sim$ 0.79. The errors in the I, Q, and U Planck maps are not propagated when calculating errors in ISP polarization. Further details of the calculation are given in \citet{2021MNRAS.502.5086C} and \citet{2023A&A...678A..63S}. Interstellar polarization is found to be 0.22\% in the V-band, which is relatively high. We subtract the ISP in the Stokes (Q, U) parameter space to correct the interstellar polarization contribution from the data. The corrected polarization degree and position angle are $P_{\mathrm{mean}}=0.35$\% with $\chi_{\mathrm{mean}}=158$ degrees over the whole wavelength range,  $P_{\mathrm{cont}}=0.34$\% and $\chi_{\mathrm{cont}}=146$ degrees in the continuum, as mentioned in the Table \ref{tab:PD_PA}. The degree of polarization after correcting for interstellar polarization has been increased twice.

\subsubsection{Line width measurement}

The black hole mass can be estimated using the Virial theorem, but line width (FWHM and line dispersion $\sigma_{\mathrm{line}}$) needs to be measured accurately. FWHM and $\sigma_{\mathrm{line}}$ have been used to estimate black hole masses, with some pros and cons in both \citep[see, e.g.,][and references therein]{2020ApJ...903..112D}. For example, the FWHM is strongly weighted by the line core and low-velocity clouds, whereas $\sigma_{\mathrm{line}}$ is weighted by the line wings and high-velocity clouds. Line FWHM is easier to measure than the $\sigma_{\mathrm{line}}$, which is affected by the S/N. This is more problematic for spectro-polarimetric data, which usually have lower S/N than flux spectra \citep{fine}. Secondly, these spectro-polarimetry data are also affected by polarization bias. According to \cite{Whittle}, inter-percentile values (IPV) are less sensitive to noise and outliers than FWHM values. IPV measurements depend on cumulative flux distribution rather than the flux density at a single point, making it more robust. Although IPV is affected by noise in the wings of lines, it is less susceptible than the line dispersion $\sigma$ \citep{fine}. 

We followed the procedure mentioned by \citet{2021MNRAS.502.5086C} for the IPV width, W. We measured the wavelengths of 25th, 50th, and 75th percentiles ($\lambda_{25}$, $\lambda_{50}$, and $\lambda_{75}$), where W is the width between $\lambda_{25}$ and $\lambda_{75}$ while $\lambda_{50}$ corresponds to the line centroid. To estimate the FWHM and IPV from the flux and polarized spectra, we first subtracted the continuum and considered the spectral range three times the FWHM of the broad H$\alpha$ components on both sides from the line center. We then estimated IPV and the FWHM using the procedure given in \citet{2004ApJ...613..682P}. We changed the line window and repeated the process until no variation was seen in IPV. Then, we computed the broadening ratio ($\delta$= W($I_p$)/W(I) ) and shift between the line centroid in I and $I_p$, mentioned in Table \ref{tab:IPW}. As the table shows, 1H0323+342 shows very slight broadening ($\approx$ 1.22), and the shift in the central line is only $\sim$211 \kms{}.

\begin{table}
\centering
\begin{tabular}{cccccc}
\hline
FWHM (I) & W (I)  & FWHM($I_p$) & W ($I_p$) & $\delta$ & $I_p$ - I shift \\
1    & 2      & 3         & 4        & 5               \\ \hline
1253 & 1941  &  1376 & 2372  & 1.22 & -211 \\
\hline
\end{tabular}
\caption{Column (1): FWHM of the $H{\alpha}$ emission line in direct light (I). (2) Inter-percentile width(between 25$^{\rm th}$ and 75$^{\rm th}$ percentile) of I. (3) FWHM of $I_p$ (4): IPW of W($I_p$). (5): Broadening ratio  (i.e W($I_p$)/W(I)). (6): shift in the line centroid of $I_p$ and I. All parameters are in units of \kms{} (except column 4).}
\label{tab:IPW}
\end{table}

\section{Modeling polarized emission using STOKES}
\label{sec:modeling}

\begin{table*}
\centering
\caption{STOKES modelling parameters}
\label{tab:stokes-model}
\resizebox{\textwidth}{!}{%
\begin{tabular}{llllllll}
\textbf{Component} &
  \textbf{Geometry} &
  \textbf{Parameter} &
  \textbf{Notation} &
  \textbf{Value} &
  \textbf{Units} &
  \textbf{Comments} &
  \textbf{Reference} \\ \hline
Continuum Source &
  Point &
  spectral index &
  $\alpha$ &
  1 &
   &
  F$_{\nu} \propto \nu^{-\alpha}$ &
   \\ \hline
BLR &
  Cylindrical &
  inner radius &
  r$_{\rm in}$ &
  0.012424 &
  pc &
  = 14.8 light days &
  Wang et al. 2016 \\
 &
   &
  outer radius &
  r$_{\rm out}$ &
  0.0712 &
  pc &
   &
  Laor \& Netzer, 1993 \\
 &
   &
  cylinder's half-height &
  a &
  0.0112 &
  pc &
  assuming half-opening angle 15 deg. and R$_{\rm mid}$ = 0.041816 pc &
   \\
 &
   &
  half-opening angle &
  $\theta$ &
  15 &
  deg &
   &
   \\
 &
   &
  cylinder's radius &
  b &
  0.02939 &
  pc &
   &
   \\
 &
   &
  line centroid &
  $\lambda_0$ &
  4861.33 &
  \AA &
  H$\beta$ &
   \\
 &
   &
  line width &
  $\Gamma$ &
  30.385 &
  \AA &
   &
   \\
 &
   &
  relative strength &
  strength &
  5 &
   &
   &
   \\
 &
   &
  Azimuthal velocity &
  v$_{\phi}$ &
  2216.85 &
  km s$^{-1}$ &
  = v$_{\rm avg}$ = $\sqrt{GM/R_{\rm mid}}$ &
   \\ \hline
Scattering region &
  Optically thick flared disk &
  inner radius &
  r$_{\rm in}$ &
  0.1117 &
  pc &
  dust reverberation mapping relation, log R = -0.89 + 0.5log(L$_{\rm V}$/10$^{44}$), here L$_{\rm V}$ = L$_{\rm 5100}$ &
  Koshida et al. 2014 \\
 &
   &
  outer radius &
  r$_{\rm out}$ &
  0.1527 &
  pc &
  \begin{tabular}[c]{@{}l@{}}the location of the outer radius of the torus is set at the location \\ where the outer radius of the BLR subtends a half-angle of 35$^{\rm o}$\end{tabular} &
  Savic et al. 2018 \\
 &
   &
  half-opening angle &
  $\theta$ &
  35 &
  deg &
   &
   \\
 &
   &
  electron density &
  n$_{\rm e}$ &
  2.53$\times$10$^5$ &
  cm$^{-3}$ &
   &
  Sniegowska et al. 2023 \\
 &
   &
  Azimuthal velocity &
  v$_{\phi}$ &
  1057.91 &
  km s$^{-1}$ &
  = v$_{\rm avg}$ = $\sqrt{GM/R_{\rm mid}^{'}}$, here R$_{\rm mid}^{'}$=$\sqrt{r_{\rm in}.r_{\rm out}}$ &
   \\ \hline
NLR &
  Double cone &
  inner radius &
  r$_{\rm in}$ &
  0.1117 &
  pc &
   &
   \\
 &
   &
  outer radius &
  r$_{\rm out}$ &
  0.1527 &
  pc &
   &
   \\
 &
   &
  half-opening angle &
  $\theta$ &
  35 &
  deg &
   &
   \\
 &
   &
  electron density &
  n$_{\rm e}$ &
  2.53$\times$10$^5$ &
  cm$^{-3}$ &
   &
   \\
 &
   &
  Azimuthal velocity &
  v$_{\phi}$ &
  1057.91 &
  km s$^{-1}$ &
   &
   \\ \hline
\end{tabular}%
}
\footnotesize{{\sc Notes.} These parameters are listed for reproducing the H$\beta$ region. Notice that the parameters for the NLR region are identical to the Scattering region - the only difference is the change in orientation of the former by 90$^{\circ}$ relative to the latter. For the H$\alpha$ region modeling, we assume similar parameters only changing the line centroid to 6562.81\AA, and the relative strength to 15.5 - this value is obtained assuming the H$\alpha$/H$\beta$ $\approx$ 3.1 \citep{Osterbrock_1984, Goodrich_1995}.}
\end{table*}

In this work, we aim to gain insights into the structure and characteristics of the scattering region surrounding the gamma-ray emitting NLS1 1H0323+342 and its polarization properties. Toward this, we employ polarization radiative transfer simulations using the {\sc STOKES} code, a state-of-the-art 3D modeling suite \citep{goosmann_gaskell_2007,marin2018,savic_etal_2018}. We use the latest version (v1.2) of the code\footnote{publicly available at \href{http://astro.u-strasbg.fr/~marin/STOKES\_web/index.html}{http://astro.u-strasbg.fr/marin/STOKES\_web/index.html}}. The code utilizes a Monte Carlo approach, tracing the path of each photon emitted from the source through the intervening medium. It accurately accounts for physical phenomena such as electron and dust scattering, tracking their interactions until photons get absorbed or reach a distant observer. Our methodology, inspired by the work of \citet{2023A&A...678A..63S}, simultaneously involves fitting the spectral regions corresponding to prominent emission lines, specifically H$\alpha$ and H$\beta$. This approach enables us to extract comprehensive information about the scattering environment and polarization properties.

\begin{figure*}
\centering
\includegraphics[width=\textwidth]{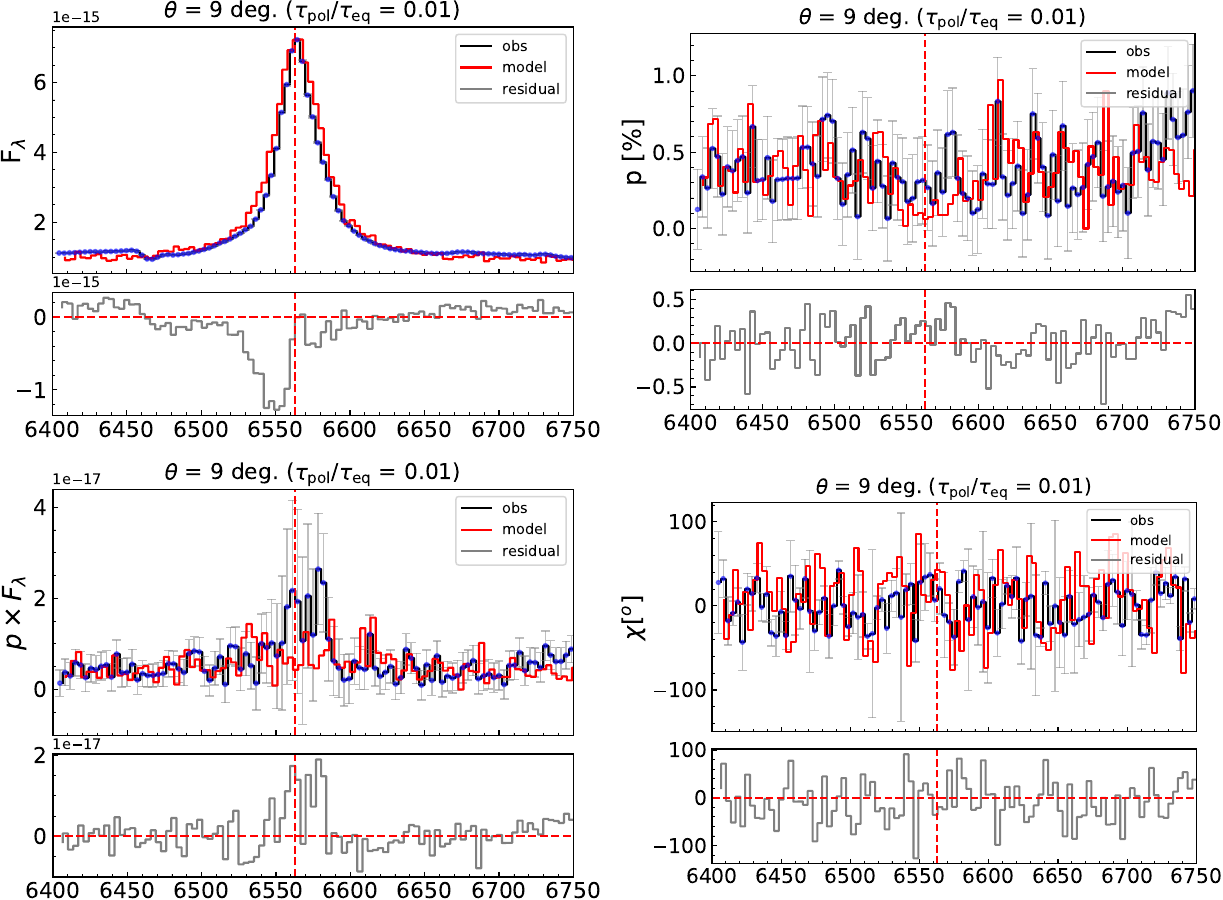}
\caption{STOKES modeling results for a representative case (viewing angle 9$^{\circ}$) for the H$\alpha$ region using un-binned spectro-polarimetric data.}
\label{Fig:fig_stokes_Ha}
\end{figure*}

\begin{figure*}
\centering
\includegraphics[width=\textwidth]{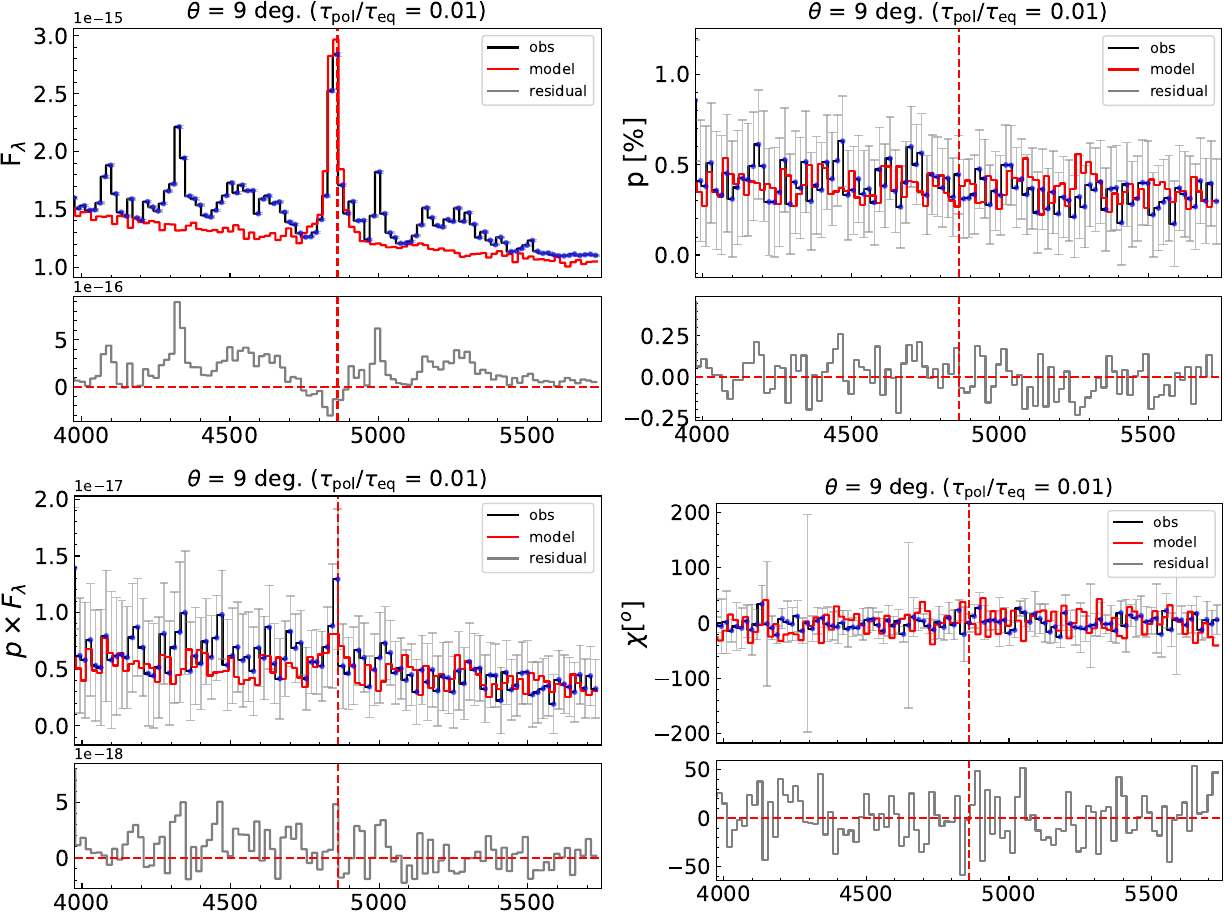}
\caption{STOKES modeling results for same case as in Figure \ref{Fig:fig_stokes_Ha} for the H$\beta$ region.}
\label{Fig:fig_stokes_Hb}
\end{figure*}

Our modeling approach begins with a central point-like continuum source emitting radiation uniformly in all directions. The emitted flux follows a power-law spectrum, with a spectral index ($\alpha$) set to 1, determined through meticulous fitting to account for both emission line and continuum contributions. The continuum windows considered in this fitting process are sufficiently wide and distant from line contamination for accuracy. This continuum source is circumscribed by a cylindrical BLR, characterized by the three spatial parameters, namely (i) the distance between the continuum source to the center of the BLR ($R_{\rm mid}$), (ii) the height of extension of the BLR above (or below) the mid-plane joining the source and the BLR ($a$), and (iii) the half-width of the BLR ($b$). These parameters are estimated using information about the BLR's inner (R$_{\rm in}$) and outer radii (R$_{\rm out}$), drawn from previous studies. The reverberation-mapped BLR size of 1H 0323+342 is estimated to be 14.8 light days or 0.012424 parsecs \citep{2016ApJ...824..149W}. We estimated the outer radius of the BLR using the analytical relation by \citet{netzer_laor_1993} for the dust sublimation radius, which has the following form: $R_{\rm out}^{\rm BLR} = 0.2L_{\rm bol, 46}^{0.5}$, where $L_{\rm bol, 46}$ is the bolometric luminosity in units of $10^{46}$ erg s$^{-1}$. The $L_{\rm bol}$ is estimated by scaling the observed monochromatic luminosity at 5100\AA~ compiled in \citet{2016ApJ...824..149W}. To complete the BLR's characterization, the velocity distribution for the H$\alpha$ line is determined by computing the average velocity along the x-y direction, assuming a Keplerian velocity distribution at a given radius from the source with a predetermined black hole mass.

In our investigation of the scattering region representing the inner edge of the torus, we modeled it as a flared disk with defined inner and outer radii and a half-opening angle. We assumed that the scattering is primarily caused by free electrons, with the number density parameter governing its properties. Similar to the approach taken for the BLR, we set the velocity distribution, assuming consistency in the inner and outer radii of the torus. The inner radius is determined using the R-L relation derived from infrared reverberation mapping studies \citep{kishimoto_etal_2007,koshida_etal_2014}. 
Following Figure 2 in \citet{savic_etal_2018}, the outer radius for the scattering region is chosen such that the BLR half-opening angle when viewed from the edge of the scattering region is 35$^{\rm o}$. This angle gives the angular extension of the BLR while keeping the observer at the outer edge of the scattering region. We choose a slightly larger value of the half-opening angle compared to \citet[][where they assumed a BLR half-opening angle = 25$^{\rm o}$]{savic_etal_2018} which allows us to reproduce well the wings of the emission lines. This value was obtained after testing with a grid of values between 0$^{\rm o}$ - 45$^{\rm o}$ (with a step size of 5$^{\rm o}$; where 0$^{\rm o}$ refers to no contribution from the BLR and 45$^{\rm o}$ means that the outer edge of the BLR coincides with the outer edge of the scattering region). Values above 45$^{\rm o}$ are discarded as they are nonphysical. Therefore, with the knowledge of the $R_{\rm out}^{\rm BLR}$ from the scaling relation of \citet{netzer_laor_1993}, and this value of the BLR half-opening angle, we can estimate the outer radius for the scattering region ($R_{\rm out}^{\rm sca}$ = $R_{\rm out}^{\rm BLR} \times {\rm sec}(\theta)$). By combining information on inner and outer radii with electron number density, we estimate the optical depth of this region. A grid of electron number density solutions is explored to identify the optimal solution aligning with spectral data gathered from observations.

In our analysis, we incorporate both equatorial and polar scatterer options, following methodologies outlined in previous studies such as \citet{2005MNRAS.359..846S} and \citet{2021MNRAS.508...79J}. The polar scatterers are positioned at the same distance from the continuum source and have identical sizes to their equatorial counterparts. We adjust the net density of the polar scatterer to achieve a satisfactory fit for both the polarized spectrum and polarization fraction, akin to the procedure employed for the equatorial scatterers. This adjustment in density influences the optical depths of the media, serving as a free parameter to optimize the STOKES parameters, including the spectrum in natural light ($F_{\lambda}$), the polarized spectrum ($p \times F_{\lambda}$), the polarization fraction ($p$) and the polarization angle ($\chi$). The modeled parameters for each component are tabulated in Table \ref{tab:stokes-model}.

We consider a grid of models assuming a range of viewing angles (i.e., 0-30 degrees) and the ratio of the optical depths of the polar region to the equatorial region ($\tau_{\rm pol}$/$\tau_{\rm eq}$, denoted as ``Factor'' as demonstrated in Figures \ref{Fig:fig_stokes_chi2_Ha} and \ref{Fig:fig_stokes_chi2_Hb}). The best-fit model to the un-binned spectro-polarimetric data with an inclination angle of 9 degrees is shown in Figures \ref{Fig:fig_stokes_Ha} and \ref{Fig:fig_stokes_Hb}. The $\chi^2$ maps for the total flux ($F_{\lambda}$), polarization fraction (p\%), polarized flux ($p \times F_{\lambda}$), and polarization angle ($\chi$) are shown in Figures \ref{Fig:fig_stokes_chi2_Ha} and \ref{Fig:fig_stokes_chi2_Hb}.

The chi-square ($\chi^2$) analysis suggests multiple solutions that are difficult to disentangle given the error bars on the observed STOKES parameters and the degeneracies in the modeled parameters. There are a few takeaways from the analysis:

\begin{enumerate}
\item The best-fit continuum slope is close to -1, one which agrees well with the H$\beta$ and the H$\alpha$ regions;
\item The best-fit results agree with the empirical H$\alpha$/H$\beta$ ``strength'' ratio (=3.1) for this source;
\item Small viewing angles are better fit to the observed distributions, i.e., 9-24 degrees;
\item A small ratio of optical depth between the polar to the equatorial scatterers is expected, i.e., a factor ($\tau_{\rm pol}$/$\tau_{\rm eq}$) = 0.05-0.1; and
\item The best-fit models reproduced well the total flux of the two Balmer lines and the depression in the degree of polarization in both cases.
\end{enumerate}

We have also considered the possibility of the two scattering regions (equatorial and polar) being made of dust instead of free electrons. We re-run the above simulation with this assumption, keeping all other model components fixed. Results from the $\chi^2$ analyses for the dust-dominated media return a valid solution only for the lowest $\tau_{\mathrm{pol}}/\tau_{\mathrm{eq}}$, i.e., 0.01. The $\chi^2$ value is quite high for higher values of this ratio. The overall shape, except the wings, of the modeled profiles for the H$\beta$ and H$\alpha$ emission lines are not reproduced in this case, and the contribution to the core of the profile is extremely low. The maximum contribution recovered using the dust-dominated media for the total flux is $\sim$1.75\% (H$\beta$) and $\sim$16.48\% (H$\alpha$). On the other hand, when the scattering media is assumed to be electron-dominated, the observed line profiles are reproduced almost perfectly: $\sim$99.27\% (H$\beta$) and $\sim$107.26\% (H$\alpha$, due to an excess in the peak flux in the modeled spectrum). This setup is also unable to recover the polarization strength (p\%) and polarization angle (PA) for the two emitting regions. Furthermore, the dust-dominated scattering media cannot reproduce the underlying continuum, suggesting that the primary scattering contributor is electron-dominated. This finding is also valid for the two emission lines (H$\beta$ and H$\alpha$). Therefore, the contributions from the dust-dominated media in this source are marginal, and the bulk of the scattering occurs closer to the inner torus edge.

We note here that these models only include the contribution of the line from the BLR. The NLR component in our {\sc STOKES} modeling acts only as a secondary scattering region in addition to the equatorial scattering region. The deficit emission noted in the {\sc STOKES} modeling comparison with observations, especially in the polarized flux ($p \times F_{\lambda}$), could be attributed to the scattering from extended regions \citep{2005MNRAS.359..846S}.

\section{Discussions}\label{sec:discussion}

We conducted spectro-polarimetric observations of the gamma-ray-emitting NLS1, 1H0323+342, renowned for its strong Fe II emission, with an R4570 of 1.14 and accretion at 37\% of the Eddington limit. The continuum polarization is modest, measuring at only 0.12\%, aligning with the observed low-polarization trend in other high-accreting NLS1 sources. \citet{1989ApJ...342..224G} performed spectro-polarimetry of 18 NLS1s, majority of which show a very low level of polarization $<0.3$\%. \citet{2011ASPC..449..431R} utilized a two-component scattering model, determining that the inclination of a line-emitting accretion disk is not a principal parameter, suggesting the distinctive properties of these objects have a physical rather than geometrical origin. Indeed, recent studies on high-accreting NLS1 show higher metallicities in these objects \citep[e.g.,][]{2019ApJ...882...79P, 2020ApJ...902...76P, 2021ApJ...910..115S, 2021A&A...650A.154P, 2022A&A...667A.105G, Floris_etal_2024}.

\subsection{Comparison with other samples}
\citet{2023A&A...678A..63S} utilized the FORS2 instrument at the \textit{Very Large Telescope} to perform spectro-polarimetric observations of three AGNs with high accretion rates. Their results showed a polarization degree of 0.15\% for IRAS 04416+1215, 0.23\% for Mrk 1044, and 0.48\% for SDSS J080101.41+184840.7. These objects are similar to our source in terms of high accretion. Another study by \citet{2021MNRAS.502.5086C} performed spectro-polarimetric observations on a sample of 25 AGNs. Their findings showed a mean polarization of 0.75\%, where a quarter of the sample had polarization levels below 0.3\%. Most of their objects had typical narrow Balmer lines with FWHM $<$2500 \kms{}. Notably, a low level of polarization is typical for Type 1 AGNs \citep[e.g.,][]{2004MNRAS.350..140S}.

\begin{figure}
    \centering
    \includegraphics[height=7cm,width=9cm]{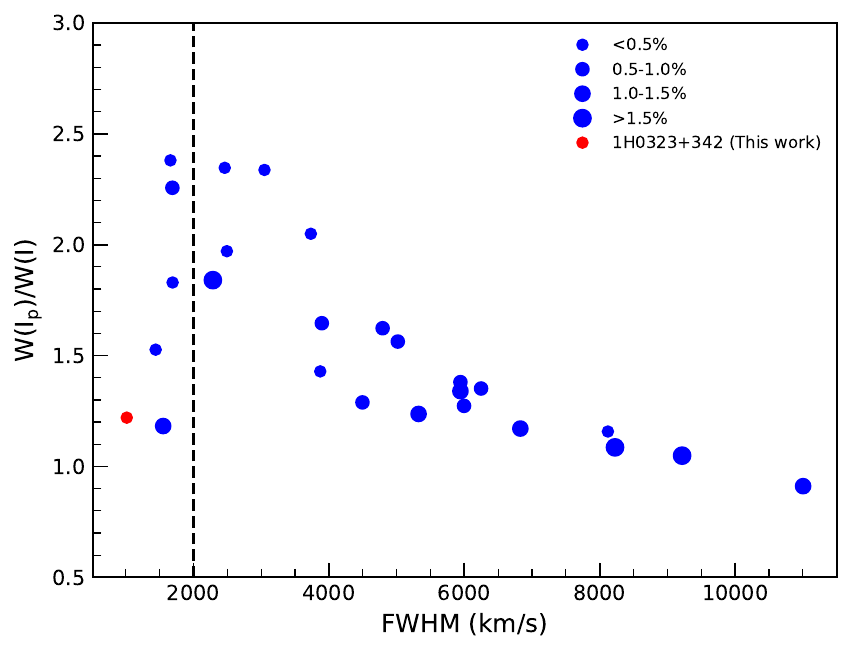}
    \caption{The polarization broadening of the BLR with respect to the $H{\alpha}$ FWHM of the sources listed in \citet{2021MNRAS.502.5086C} along with our source 1H0323+342 (red). The different sizes of the markers are associated with their continuum polarization value. The vertical line at FWHM of 2000 \kms{} shows the FWHM criteria used to define NLS1s.}
    \label{fig:wipwifwhm}
\end{figure}

We plotted the H$\alpha$ broadening ratio of polarized flux to total flux for our source in Figure \ref{fig:wipwifwhm} along with the sources presented in \citet{2021MNRAS.502.5086C} accounting for the continuum polarization value. Most of the sources of \citet{2021MNRAS.502.5086C} with continuum polarization larger than 1\%  have a decreasing trend of $\delta$ with increasing line width, i.e., the ratio of the line width between polarized spectra to flux spectra decreases. However, sources with low polarization $<1$\% do not show any clear pattern with line width.  1H0323+342 is situated in the lower-left corner, having the lowest FWHM among all the sources in the Figure. The continuum polarization of 1H0323+342 is also very low, $<1$\%, and the broadening is only 1.22. Although the high FWHM objects show negligible broadening between polarized and total flux spectra, the situation is complicated for objects with FWHM lower than 2000 \kms{}. Here, objects with different polarization labels are mixed, and broadening by a factor of 2.5 can be seen.

\subsection{Polarization due to non-thermal emission?}

\begin{figure}
\centering
\includegraphics[height=6cm,width=9cm]{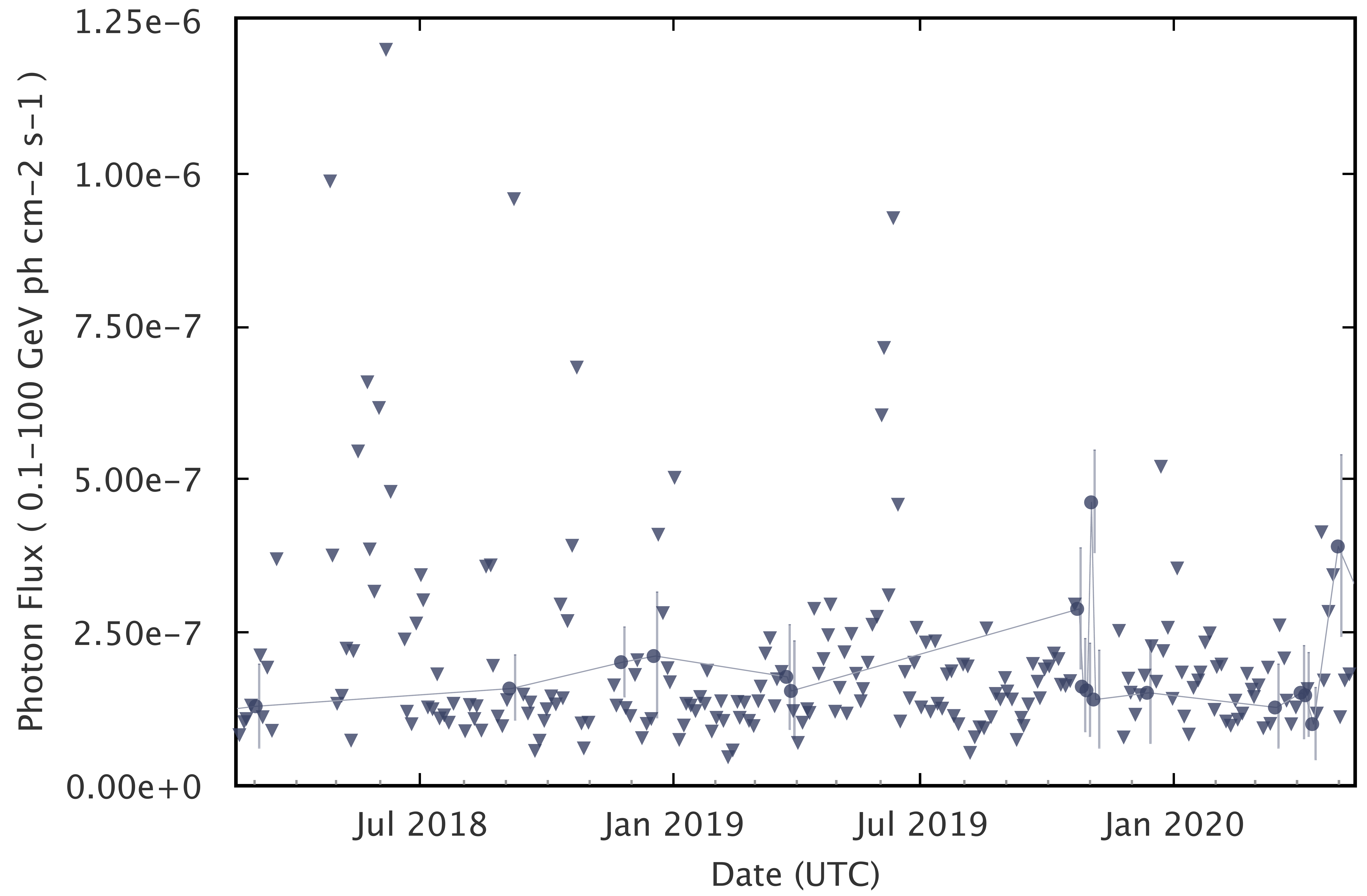}
\caption{\textit{Fermi} lightcurve of 1H0323+342 obtained through \textit{Fermi} All-sky Variability Analysis (FAVA) - Light Curve Generator. The source was in the quiescent phase during our spectro-polarimetric observation on 20$^{\rm th}$ Dec. 2019. The circles denote the detection, while the downward triangles denote the upper limit of the photon flux.}
\label{Fig:fig_fermi}
\end{figure}

1H0323+342 is a well-known gamma-ray emitting NLS1 detected by \textit{Fermi}-LAT in the energy range of MeV-GeV. Its spectral-energy distribution shows the signature of non-thermal contribution in the optical and UV but is dominated by accretion disk radiation \citep[see][]{Paliya_2019}. Previous SPOL imaging polarimetric observations of 1H0323+342 have shown its polarization degree increasing from 0.55 on 05/09/2013 to 1.40 on 06/09/2013 and then decreasing to 0.13 on 05/10/2013\footnote{\url{http://james.as.arizona.edu/~psmith/Fermi/DATA/Objects/1h0323.html}}. \citet{2014PASJ...66..108I} performed high cadence polarization monitoring of 1H0323+342 around July 2013 when the source was in a high gamma-ray activity state and found the polarization degree of the source increased from 0-1\% in quiescence (before July) to $\sim3$\% at maximum (in July) and then declined to the quiescent level. Overall, this high state lasted for 20 days, showing drastic changes in polarization. Therefore, a high degree of polarization from non-thermal emissions is expected during a high activity state. The archival \textit{Fermi} lightcurve shows the source was in a quiescence state (see Figure \ref{Fig:fig_fermi}) during our observations with a photon flux (upper-limit) of 1.37 $\times 10^7$ ph cm$^{-2}$ s$^{-1}$ in the energy range of 0.1 to 100 GeV\footnote{\url{https://fermi.gsfc.nasa.gov/ssc/data/access/lat/LightCurveRepository/source.html?source_name=4FGL_J0324.8+3412}} \citep{2023ApJS..265...31A}. Therefore, the continuum polarization of our source during our observations is not due to non-thermal emission from the jet.

\subsection{Dilution from host-galaxy starlight}

The host galaxy can dilute the intrinsic polarization of the source. 1H0323+342 shows a peculiar low-surface brightness feature in the HST images that has been suggested as a one-armed spiral host structure \citep{2007ApJ...658L..13Z} or a ring-like structure in the optical B/R-band images \citep{2008A&A...490..583A}. \citet{2014ApJ...795...58L} performed 2D image decomposition, and their Fourier spectrum shows a ring-like structure. The optical and IR images are well-fitted by an AGN and a Sersic bulge of index n=2.8. The source is low-luminous at z =0.062. Therefore, the host-galaxy contribution must be present in our observed spectrum; however, the optical spectrum with a 1.5 arcsec aperture does not show any absorption feature to decompose the host contribution. Rather, the spectrum is blue with a slope of -1.43, suggesting a strong AGN signature in the small spectroscopic aperture. The host contributes $\sim$40\% to the total flux estimation based on the HST image by \citet{2007ApJ...658L..13Z}, a similar value is obtained using the relationship given by \citet{2023MNRAS.521L..11J} based on SDSS quasars spectra. This contribution would be significantly smaller in our 1.5 arcsec aperture than the SDSS fiber (3 arcsec in diameter). Hence, the host contribution is expected to be smaller in our spectro-polarimetric data.

\subsection{Inferences about the inclination and geometry}\label{sec:geometry}

Despite having some inter-dependency among the fitted parameters in STOKES modeling (see section \ref{sec:modeling}) and a very low-polarization signature, we cautiously inferred a small viewing angle of 9-24 degrees for this source. A small viewing angle is expected as the source is a well-known gamma-ray-detected NLS1 with Doppler-boosted emissions detected by \textit{Fermi}. Previous authors have constrained the viewing angle to be small $\sim10$ degrees \citep{2016RAA....16..176F}. By modeling the 15 GHz lightcurve, \citet{2018ApJ...866..137L} estimated the jet viewing angle of $9.06^{+1.04}_{-8.91}$ deg. 

Inference can be drawn about the geometry of the BLR from the ratio of the width of the H$\alpha$ emission lines in the polarized spectrum to the total flux spectrum. Assuming the $V_{\mathrm{kep}}$ is the width of the H$\alpha$ line in the polarized spectrum, and the $V_{\mathrm{obs}}$ is the observed width of H$\alpha$ in the total flux spectrum, which is affected by the geometry of the BLR, the $V_{\mathrm{obs}}$ according to \citet{2006A&A...456...75C} can be written as 
\begin{equation}
    V_{\mathrm{obs}} = V_{\mathrm{kep}} \sqrt{ (H/R)^2 + \sin^2 i }
\end{equation}
and the virial factor is 
\begin{equation}
    f =   [ (H/R)^2 + \sin^2 i ]^{-1}
\end{equation}

Where $H$ is the disk thickness at the radius $R$ from the center, it has been discussed in detail in \citet{2006A&A...456...75C} that the BLR clouds must be able to capture a significant fraction of its ionizing photons, and therefore, it can not be a thin disk; rather, it should have vertical thickness supported by the turbulence motion within the BLR clouds. 1H0323+342 is a high-accreting NLS1 whose H$\alpha$ lines are well-fitted with two Gaussian/Lorentzian components having an FWHM/$\sigma$ ratio of 1.4. The turbulence, which has isotropic velocity distribution, affects the line wings and increases $\sigma$ compared to the FWHM, which is dominated by the line core. 

The inter-percentile width mentioned in Table \ref{tab:IPW} can be used to calculate the inclination. The $V_{\mathrm{obs}}/V_{\mathrm{kep}}=0.82$ implies $f =1.48$, and the disk thickness must be high enough to explain the low-inclination angle inferred from the STOKES modeling. Assuming the inclination angle $i=9$ degrees implies a $H/R$ = 0.8 or disk half-opening angle of $\sim$40 degrees ($\omega = \arctan [H/R]$). A relatively higher inclination angle of 24 degrees will correspond to a H/R = 0.7 and, consequently, a disk half-opening angle of 35 degrees. Indeed, recent dynamical modeling of reverberation mapping data, e.g., \citet{2018ApJ...866...75W}, supports the argument that BLR is a thick disk where many objects show a half-opening angle ($\omega$) larger than 40 degrees (here $\omega$ =90\degree corresponds to a sphere). The BLR can have flared disk-like geometry where $H$ increases with R as $H\propto R^{\alpha}$ \citep[see discussion in][]{2006A&A...456...75C}. Such flaring `BOWL'-shaped BLR models can explain various observational signatures, including the mechanism responsible for producing Lorentzian profiles in Balmer lines in low-inclination systems dominated by turbulence \citep[see][]{2012MNRAS.426.3086G}.

\section{Summary}\label{sec:summary}
We performed spectro-polarimetric observations of gamma-ray emitting NLS1 1H0323+342 using the SPOL instrument mounted at \textit{MMT}. The degree of polarization in the continuum is 0.122 $\pm$ 0.040\%, while in the line, it is 0.265  $\pm$ 0.280\%. The total flux spectrum shows the presence of strong H$\alpha$ and H$\beta$ emission lines with FWHM of H$\alpha$ of 1015 \kms{} and FWHM of H$\beta$ of 1554 \kms{}, providing an estimate of a black hole mass of $10^{7.24\pm0.01} M_{\odot}$. The polarized spectrum shows a similar line width as the flux spectrum. Radiative transfer simulations using STOKES code assuming a point-like continuum source, a cylindrical BLR, and an optically thick flared disk-like scattering region can reproduce the total flux of the H$\alpha$ and H$\beta$ lines along with the depression in the polarization degree. The bulk of the scattering occurs closer to the inner edge of the torus dominated by electrons, primarily arising from the equatorial region that accounts for $\gtrsim$90\% of the polarization in both line and continuum in this source. A small viewing angle of 9-24 degrees is favored to fit the observed data, consistent with 1H0323+342 being a gamma-ray detected NLS1 where the jet is pointing towards us. The low-inclination angle of $i=9$ degree obtained from the STOKES modeling implies a $H/R$ = 0.8 or half-opening angle $\sim$40 degrees, leading to a negligible broadening of the polarized spectrum compared to the flux spectrum even though viewed nearly face-on. 
\label{s:discussion}

\section*{Data availability}
The data underlying this article will be shared on reasonable request to the corresponding author.

\section*{Acknowledgement}
We thank the referee for her/his valuable comments and suggestions, which helped to improve the quality of the manuscript. The authors thank Paul Smith for his valuable support in the observations and SPOL data reduction and Alessandro Capetti for his help in the estimation of inter-stellar polarization. JJ and SR acknowledge the partial support of SERB-DST, New Delhi, through SRG grant no. SRG/2021/001334. SP acknowledges the financial support of the Conselho Nacional de Desenvolvimento Científico e Tecnológico (CNPq) Fellowships 300936/2023-0 and 301628/2024-6. SP is supported by the international Gemini Observatory, a program of NSF NOIRLab, which is managed by the Association of Universities for Research in Astronomy (AURA) under a cooperative agreement with the U.S. National Science Foundation, on behalf of the Gemini partnership of Argentina, Brazil, Canada, Chile, the Republic of Korea, and the United States of America. NS acknowledges the financial support provided under the National Post-doctoral Fellowship (NPDF; Sanction Number: PDF/2022/001040) by the Science \& Engineering Research Board (SERB), a statutory body of the Department of Science \& Technology (DST), Government of India. Observations reported here were obtained at the MMT Observatory, a joint facility of the University of Arizona and the Smithsonian Institution.

\bibliography{references}

\begin{appendix}

\section{Supplementary figures}

\begin{figure*}
\centering
\includegraphics[width=0.8\textwidth]{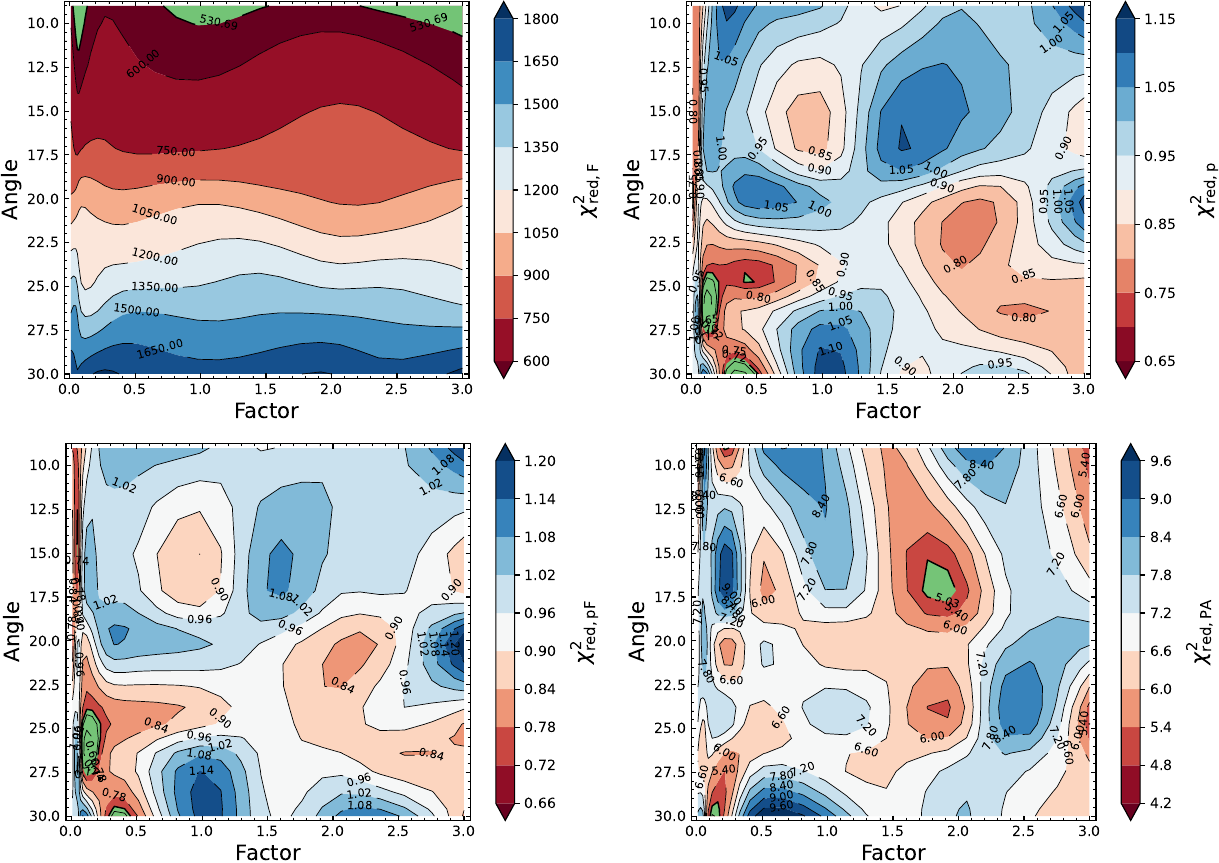}
\caption{$\chi^2$ maps for the four STOKES parameters for the H$\alpha$ region. Here, the ratio of the optical depths of the polar region to the equatorial region ($\tau_{\rm pol}$/$\tau_{\rm eq}$, denoted as ``Factor'' and inclination/viewing angle is refereed as ``Angle''. The shaded region in green marks the region within 10\% of the global minimum $\chi^2$ value.}
\label{Fig:fig_stokes_chi2_Ha}
\end{figure*}

\begin{figure*}
\centering
\includegraphics[width=0.8\textwidth]{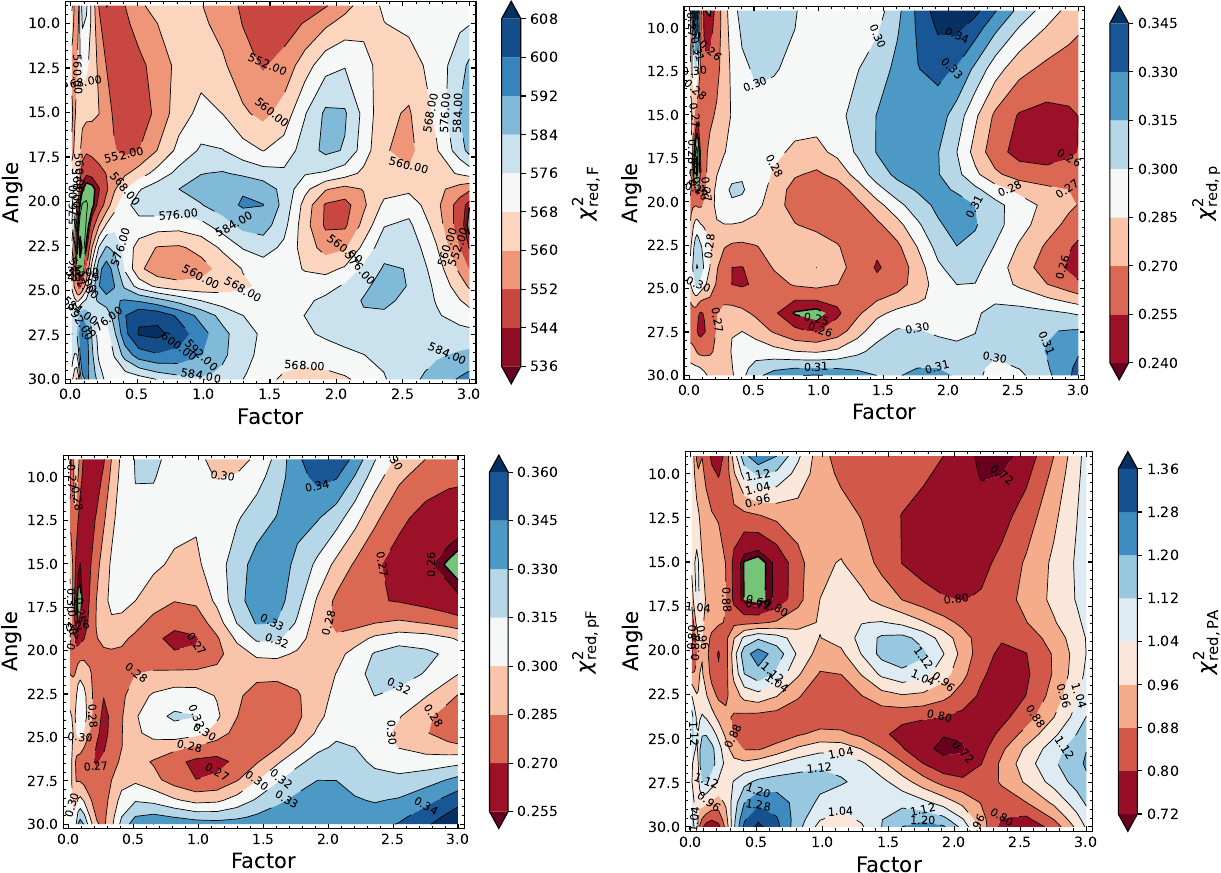}
\caption{$\chi^2$ maps for the four STOKES parameters for the H$\beta$ region.}
\label{Fig:fig_stokes_chi2_Hb}
\end{figure*}

\end{appendix}

\end{document}